\documentclass[12pt]{article}

\usepackage[margin=1in]{geometry}
\usepackage{setspace}
\usepackage{amsmath,amssymb,amsthm,mathtools}
\usepackage{graphicx}
\usepackage{microtype}
\usepackage{placeins}
\usepackage{enumitem}
\usepackage{tikz}
\usepackage{pgfplots}
\usepackage{subcaption}
\usepackage{booktabs,tabularx,threeparttable}

\usepackage[authoryear,round]{natbib}
\usepackage[hidelinks,
 bookmarksopen=true,
 bookmarksopenlevel=2,
 bookmarksnumbered=true,
 pdfpagemode=UseOutlines]{hyperref}
\usepackage{bookmark} % improves and stabilizes bookmarks



\newcommand{\Prb}{\mathbb{P}}
\newcommand{\E}{\mathbb{E}}
\newcommand{\cFB}{c^{\mathrm{FB}}}
\newcommand{\cLF}{c^{0}}
\newcommand{\ubar}{\underline{c}}
\newcommand{\obar}{\overline{c}}

\newtheorem{assumption}{Assumption}
\newtheorem{theorem}{Theorem}
\newtheorem{lemma}{Lemma}
\newtheorem{proposition}{Proposition}
\newtheorem{corollary}{Corollary}
\newtheorem{definition}{Definition}
\theoremstyle{remark}
\newtheorem{remark}{Remark}

\title{Paying for Failure in Expert Advice\thanks{An earlier version of this article circulated under the title ``Risky Advice and Reputational Bias.'' We are grateful to Costas Cavounidis, Olivier Gossner, Peter Hammond, Vitalijs Jascisens, Margarita Kirneva, Steven Kivinen, Olga Kuzina, Harry Pei, Alessandro Riboni, Ludvig Sinander, Bruno Strulovici, Stepan Svistunov, Nikhil Vellodi, Alexey Verenikin, and all the participants of the 2022 Transatlantic Theory Workshop for useful comments and discussions. This research was supported by the French National Research Agency (ANR) through the ``Investissements d'Avenir'' program (ANR-17-EURE-0010). All remaining errors are ours.}}
\author{%
 Georgy Lukyanov\thanks{Department of Economics, \emph{Toulouse School of Economics}. Email: \texttt{georgy.lukyanov@tse-fr.eu@}.} \and
 Anna Vlasova\thanks{Department of Economics, \emph{CREST—École Polytechnique}. Email: \texttt{anna.vlasova@polytechnique.edu}.} \and
 Maria Ziskelevich\thanks{International College of Economics and Finance, \emph{National Research University—Higher School of Economics}. Email: \texttt{mmziskelevich@edu.hse.ru}.}%
}
\date{}

\begin{document}
\maketitle

\begin{abstract}
A failed recommendation is visible; an unproposed project is not. How should an organization pay an adviser whose private confidence determines which projects reach the margin? We show that the least-cost instrument is protection after failure rather than a success bonus. Because risky advice selects the upper tail of confidence, success pay leaks to recommendations that would occur anyway, while failure protection is concentrated at the margin. We establish unique nonpooling implementation, show that full correction is never optimal, and, while advice still needs encouragement, find confidential internal review substitutes for explicit career insurance more effectively than transparent review.
\end{abstract}


% =========================================================
% ===================== Main Text =========================
% =========================================================

\section{Introduction}\label{sec:intro}

An investment officer decides whether to take an acquisition to the
committee; an engineer decides whether to certify a technology that has never
been run at scale; a product team decides whether to advise a launch. In each
case, the paper studies the possibility that the organization does not observe
what the specialist observed---it sees the recommendation and, some time
later, whether the project worked. That asymmetry is the source of the
problem studied here. A recommendation that fails creates an observable
record with a name attached to it, whereas the value forgone when a project
is never proposed is not observed.\footnote{The
asymmetry is one of \emph{observability}, not of preferences. Nothing in what
follows requires the adviser to be biased, risk averse, or empire building.
It is enough that the outside assessment of her ability responds to the
outcomes of the projects she puts forward, and that it cannot respond to the
projects she quietly declines to put forward.} An adviser who fully shares the
organization's objectives, and who suffers from no distortion other than not
wanting to be thought incompetent, will therefore still decline projects that
the organization would want approved.

The question we ask is a design question: how should an organization
compensate a career-concerned adviser whose private confidence determines
which projects reach the decision margin? A natural response is to pay
for success---to attach a bonus to the bold project that works. It is shown
here that success pay is the expensive instrument, and that the cheapest
contract does something that sounds perverse when stated baldly: it pays the
adviser \emph{after a failure}. This result is not an artifact of risk
aversion, and it is worth flagging as counterintuitive, since common sense
would suggest that rewarding failure is precisely the way to get bad projects
recommended. Neither reading is right. The payment is not a reward for
error; it is insurance against the reputational consequences of a
recommendation that was reasonable when it was made and unlucky
afterwards.\footnote{In practice the instrument need not look like a cash
transfer at all. A retention guarantee, a protected evaluation window,
severance protection, a commitment not to reduce responsibilities because an
approved project failed, or an evaluation rule that records whether a decision
was defensible \emph{ex ante} all deliver the same marginal incentive as the
failure payment $t_-$ in the model. What matters is only that the
organization can commit to it and that the adviser values it.}

The intended organizational setting has four features: the adviser privately
observes her confidence in the project; the organization can commit to an
implementation rule; the realized outcome feeds an internal or external
assessment of the adviser's ability; and the adviser is not the residual
claimant on project value. Internal capital allocation, R\&D stage gates, and
product-launch approvals can have this shape. The fourth feature is the one
that creates room for corrective contracting: because the organization retains
some surplus from marginal projects, it can trade that surplus off against the
cost of changing the advice threshold.

The mechanism behind everything that follows is selection. An adviser
recommends the risky action only when her posterior confidence clears a
threshold, so, conditional on risky advice, average confidence is strictly
above the confidence of the marginal adviser---the only adviser whose behavior
the contract is trying to change. A success bonus is therefore paid mainly on
high-confidence, inframarginal recommendations that would have been made in
any case. Failure protection, by contrast, is paid disproportionately on the
lower-confidence recommendations near the margin, because those are the ones
that fail. Per unit of incentive delivered where it is needed, failure
protection is strictly cheaper. The comparison is a statement about where a
payment lands, not about how the adviser feels about
risk.\footnote{Consequently the result does not require risk aversion; we
assume risk neutrality throughout. The standard insurance argument for
cushioning failure and the selection argument given here are logically
independent, and the paper isolates the second.}

We formalize the argument in a delegated-assessment model. Ability is high or
low and determines signal precision, but it is initially unknown both to the
adviser and to the market; the adviser observes a continuous private signal
about project quality and recommends a safe or a risky action. The
organization has committed to implement risky advice, and it can condition
nonnegative incremental compensation on safe advice, on a successful risky
project, or on a failed risky project. The market observes the same three
histories and updates its belief about ability. The adviser holds an
outcome-aligned material stake in the project alongside a linear career
payoff.

Our first result identifies the reputational distortion from primitives rather
than assuming it. We use a tractable continuous-signal family in which high
ability is a Blackwell-more-informative experiment. Successful risky advice
raises reputation, failed risky advice lowers it, and safe advice leaves
reputation at the prior. At any interior cutoff, the expected reputation
generated by risky advice at the margin is strictly below the reputation
generated by safety, so career concerns push the laissez-faire threshold above
the efficient one. It is the asymmetry of \emph{informativeness}, not any
asymmetry of the payoff function, that produces the wedge: safe advice is
uninformative about ability, and an uninformative history is worth the prior,
which the marginal risky recommendation cannot match.

Our central theorem is considerably more general than that signal family. Fix
any interior confidence threshold that requires additional encouragement.
Among all nonnegative transfers that support it, the unique least-cost
incremental contract pays only after failure: if $D(c)$ is the marginal
incentive deficit at target $c$, then
\[
 t_-^*(c)=\frac{D(c)}{1-c},\qquad
 t_+^*(c)=t_S^*(c)=0.
\]
The strict cost comparison rests on nothing more than the upper-tail selection
identity $\E[p\mid p\geq c]>c$, which holds for any atomless nondegenerate
distribution of confidence. It does not rely on risk aversion, on dynamic
learning, or on privately known ability.\footnote{The last of these matters
for interpretation. \citet{Chen2015} shows that if the adviser privately
knows her ability the sign of the distortion can reverse and \emph{over}
investment in risky projects becomes the equilibrium outcome. Our benchmark
deliberately keeps ability commonly unknown so that the reputational wedge has
an unambiguous sign and the contracting problem is well posed; Section
\ref{sec:conclusion} returns to what changes when that assumption is dropped.}

Supporting a cutoff under the beliefs it generates is not yet an
implementation theorem, and we do not treat it as one. We therefore define
the adviser's strategy over the full set of measurable, possibly mixed
recommendation rules. A primitive success-dominance condition implies that
any nonpooling equilibrium must be an upper cutoff, and the
failure-protection schedule is strictly decreasing in the target cutoff, which
delivers a unique nonpooling equilibrium. Full uniqueness additionally
requires an explicit upper-tail-consistent restriction on beliefs at pooling
histories. We state that refinement rather than hide the dependence of a
pooling claim on off-path beliefs: outcome histories are simply unreached when
the adviser never recommends risk, and no appeal to Bayes' rule alone can
settle what the market would then conclude.

The organization then chooses the advice threshold as well as the contract.
It never induces advice below the first-best threshold and never chooses a
threshold above laissez faire. More sharply---and this is the result we would
single out for practice---exact first-best implementation is never optimal. At
the first-best threshold the marginal project-surplus loss from a small
increase in caution is zero, while the compensation saving is first order, so
every optimum either leaves laissez faire untouched or corrects the
reputational distortion only partially. In the uncapped canonical model the
curvature needed for uniqueness follows from primitives whenever failure is at
least as costly to the organization as success is valuable; for a binding cap
or a noncanonical confidence distribution, a transparent marginal
single-crossing condition delivers the same unique threshold and the same
endpoint intervention test.

Two organizational constraints sharpen the picture. First, downside protection
may be capped---by governance rules, by internal pay equity, or simply by the
difficulty of explaining to anyone why money changed hands after a project
failed. The least-cost contract then exhausts failure protection up to the cap
and uses success pay only for the incentive that remains, and relaxing the cap
weakly lowers the optimal threshold. Second, the organization may review risky
recommendations before implementation. A one-sided review detects bad projects
and vetoes them. If the market observes the veto, review prevents losses but
does not touch the pure career wedge, for the revealing reason that a veto and
a failed launch say exactly the same thing about ability. If instead the veto
is confidential and pooled with safe advice, review shields the adviser from
part of the adverse updating, and confidential review therefore reduces the
explicit compensation needed to support a fixed target by more than transparent
review does. The comparison is deliberately local to the region in which the
organization is still trying to encourage risk; sufficiently strong review can
reverse the underlying distortion, and past that point the ranking need not
survive.\footnote{This is not a defense of institutional secrecy in general.
The committee literature contains examples in both directions---see
\citet{Levy2007} on how opaque procedures can push members toward pre-existing
biases, and \citet{GersbachHahn2012} on transparency and the incentive to
acquire information---and our statement is narrower than either: while career
concerns are suppressing desirable risk, confidentiality supplies insurance
that transparency does not.}

\subsection*{Related literature}

The paper contributes to three literatures. First, work on career concerns
shows how inferred ability affects effort, project choice, communication, and
the interaction between implicit and explicit incentives
\citep{Holmstrom1999,GibbonsMurphy1992,OttavianiSorensen2006,
SuurmondSwankVisser2004,Chen2015,LossRenucci2021}. The closest contracting
paper is \citet{CaruanaCelentani2002}, who show in a two-period managerial
investment model with privately informed managers that generous compensation
following poor results may offset career distortions. Our distinct object is
the extensive margin of advice, selected on private confidence; that selection
is what delivers a strict comparison between the cost of success and failure
payments, and what links the least-cost contract to an endogenously chosen
recommendation threshold. A recent working paper by \citet{GanLi2026} instead
studies robust effort implementation when effort changes the informativeness of
performance and equilibrium beliefs about effort can be pessimistic. We
remove hidden effort and strategic uncertainty in order to isolate the
private-confidence margin.\footnote{The two exercises are complements rather
than rivals. In \citet{GanLi2026} the difficulty is that the market's
conjecture about effort is itself in play, so the employer must implement in
\emph{every} equilibrium; here the adviser's action is a recommendation rather
than an effort, the informativeness of the public record is fixed by ability,
and the difficulty is instead that the population being paid is endogenously
selected.}

Second, tolerance for failure appears in innovation, experimentation, and
project-finance contracts. \citet{Manso2011} emphasizes tolerance for early
failure together with rewards for long-run success; \citet{HalacKartikLiu2016}
study long-term experimentation with adverse selection, moral hazard, and
private learning; and \citet{KhalilLawarreeRodivilov2020} obtain rewards after
failed experimentation from screening in a model combining experimentation and
production. Especially close is \citet{MeunierPonssard2024}: in a
green-project subsidy model, failure-only support limits windfall payments to
high-success-probability projects that would have invested without support.
Neither pay following failure nor its selection-based cost advantage is
therefore new in isolation, and we do not claim otherwise. The present paper
tries to fill in a different gap---to derive the force from endogenous career
evaluation in delegated advice, and then to solve the implementation and
organizational-design problems that go with it: unrestricted nonpooling
implementation, the principal's second-best advice threshold, a cap-induced mix
of failure and success pay, and the disclosure design of internal review. The
model is static and contains neither an experimentation path nor hidden
effort.\footnote{The dynamic counterpart is a different animal.
\citet{HalacKremer2020} show that when a manager learns privately over time and
recognizing failure is itself reputationally costly, the distortion is delay
rather than excessive caution at the outset, and it can grow with expected
project quality. Our static benchmark has no continuation decision to
distort, which is exactly why the extensive margin of advice can be isolated.}

Third, research on advice, committees, and disclosure studies how reputational
motives interact with preselection and transparency
\citep{BourjadeJullien2011,SongThakor2006,SchulteFelgenhauer2017,
Levy2007,BagSharma2019,LiuSanyal2012}. Relatedly,
\citet{LiuSanyal2012} introduce a second opinion that may reverse the
principal's initial action under career concerns. We instead treat the
disclosure of an internal veto as part of an incentive system: it changes the
outside inference attached to the veto and therefore the compensation cost of
implementing advice. This connects information architecture to explicit
career insurance.\footnote{Two
qualifications on how our exercise sits relative to this literature. First, we
do not model the adviser's information acquisition, which is the margin
\citet{BarIsaac2012} shows can be either helped or hurt by making project
choice observable; in our model precision is a fixed function of ability.
Second, our adviser is unbiased by construction, so the paternalistic and
persuasion distortions studied for biased experts---see, e.g.,
\citet{Lightle2014}---are absent, and every distortion that appears is
reputational.}

The rest of the paper proceeds as follows. Section~\ref{sec:model} presents
the model. Section~\ref{sec:reputation} derives reputational conservatism and
the laissez-faire cutoff. Section~\ref{sec:contract} establishes the
least-cost contract and the implementation result.
Section~\ref{sec:cap} studies a cap on downside protection.
Section~\ref{sec:principal} characterizes the organization's target.
Section~\ref{sec:review} analyzes internal review and disclosure.
Section~\ref{sec:numerical} provides a numerical illustration, and
Section~\ref{sec:conclusion} concludes. Proofs and auxiliary results are
collected in the Appendix.

\section{Model}\label{sec:model}

The model is built to be as thin as the argument allows. Ability is binary
and unknown to everyone, including the adviser, so that reputational updating
has an unambiguous direction; the private signal is continuous, so that the
recommendation margin is a genuine margin rather than a choice among two or
three types; and the adviser's material stake is proportional to the
organization's, so that any distortion which survives is reputational by
construction. Each of these is doing a specific job, and we flag below where
relaxing it would change a conclusion.

\subsection{Information}

Project quality is \(\omega\in\{1,0\}\), with equal prior probabilities.
Ability is \(\theta\in\{H,L\}\), independent of project quality, with
\(\Prb(\theta=H)=\rho\in(0,1)\). Ability is initially unknown to both the
adviser and the market. Conditional on \((\theta,\omega)\), the adviser
observes \(x\in[0,1]\) with density
\begin{align}
 f_\theta(x\mid 1)&=1+a_\theta(2x-1),&
 f_\theta(x\mid 0)&=1-a_\theta(2x-1), \label{eq:density}
\end{align}
where
\[
 0\leq l\equiv a_L<h\equiv a_H<1,\qquad
 b\equiv \rho h+(1-\rho)l\in(0,1).
\]
The densities are strictly positive and integrate to one. The
state-likelihood ratio is increasing in \(x\), and the low-ability
experiment is a Blackwell garbling of the high-ability
experiment.\footnote{The garbling is explicit and worth stating, because it is
what makes ``high ability'' mean something in this model: a low-ability
adviser's signal can be generated from a high-ability adviser's signal by
retaining it with probability \(l/h\) and otherwise replacing it with an
independent uniform draw. Higher ability is therefore more information in the
Blackwell sense, and not merely a different information structure.} The
adviser's posterior probability of project success is
\begin{equation}
 p(x)\equiv\Prb(\omega=1\mid x)
 =\frac{1-b}{2}+bx. \label{eq:posterior}
\end{equation}
Thus \(p\) is strictly increasing in \(x\) and is uniformly distributed on
\[
 [\ubar,\obar]
 =
 \left[\frac{1-b}{2},\frac{1+b}{2}\right].
\]

\subsection{Advice, outcomes, and payoffs}

The adviser recommends the safe action \(S\) or the risky action \(R\).
The organization has committed to implement \(R\). If \(\omega=1\), the
risky project succeeds and produces \(G>0\) for the organization; if
\(\omega=0\), it fails and produces \(-L<0\). Safety yields zero. The
first-best confidence threshold is
\begin{equation}
 \cFB=\frac{L}{G+L}. \label{eq:firstbest}
\end{equation}
We assume it is interior:
\begin{equation}
 \ubar<\cFB<\obar. \label{eq:interior}
\end{equation}

The adviser has a limited but aligned stake in performance. She receives
\(g>0\) after success, \(-\ell<0\) after failure, and zero after safety,
where
\begin{equation}
 \frac{\ell}{g+\ell}=\frac{L}{G+L}=\cFB. \label{eq:alignment}
\end{equation}
Absent career concerns, the adviser and organization therefore agree about
which projects should be implemented.

The market observes one of three public histories:
\[
 h\in\{S,+,-\},
\]
where \(+\) and \(-\) denote a successful and failed risky recommendation.
Let \(\mu_h\) be the market's posterior probability that ability is high
after history \(h\). The adviser's continuation career value is
\[
 V(\mu_h)=\gamma\mu_h,\qquad \gamma>0.
\]

The organization offers nonnegative incremental payments
\[
 (t_S,t_+,t_-)\geq0.
\]
These payments are made above an immutable base salary. The adviser is an
incumbent, and participation is slack for the incremental contract. This
normalization isolates the cost of changing behavior. If a freely
adjustable base wage instead makes participation bind, failure protection
remains weakly cost minimizing, but strict comparisons of total wage bills
can disappear.\footnote{The reason is worth being explicit about, since the
whole paper is a cost comparison. With a fully flexible base wage the
employer can claw back in the wage whatever it hands over in the outcome
schedule, so any two schedules delivering the same expected utility to the
adviser cost the same in total. What survives is the ranking of
\emph{incremental} instruments, which is the object an organization actually
chooses when it decides how to treat a failed project.}

The word ``advice'' denotes a delegated screening decision, not cheap talk.
Before the signal is realized, the organization commits that an \(R\)
recommendation clears the project for implementation. Standing investment
committee rules, preauthorized R\&D gates, and delegated product-launch
authority have this form. If the organization could observe \(x\), or
could freely override the adviser after learning the recommendation, the
relevant problem would instead be one of information acquisition or
strategic communication; neither friction is claimed here.

\subsection{Timing and equilibrium}

The timing is:
\begin{enumerate}[label=(\roman*)]
 \item the organization offers the incremental contract;
 \item the adviser observes \(x\) and recommends \(S\) or \(R\);
 \item risky advice is implemented and the outcome is realized;
 \item the transfer is paid and the market updates the adviser's ability.
\end{enumerate}

An unrestricted behavioral advice strategy is a measurable function
\(r:[0,1]\to[0,1]\), where \(r(x)\) is the probability of risky advice.
An equilibrium consists of such a strategy and market beliefs satisfying
sequential rationality and Bayes' rule at on-path histories. A strategy is
nonpooling if both recommendations occur with positive probability.

For later use, write \(z\in(0,1)\) for a raw-signal cutoff and
\[
 c=p(z)=\ubar+bz
\]
for the corresponding confidence cutoff. Under the rule
\(R\Longleftrightarrow x\geq z\), the history probabilities and ability
posteriors are
\begin{align}
 P_+(z)&=\frac12(1-z)(1+bz),&
 P_-(z)&=\frac12(1-z)(1-bz),&
 P_S(z)&=z, \label{eq:historyprob}\\
 \mu_+(z)&=\frac{\rho(1+hz)}{1+bz},&
 \mu_-(z)&=\frac{\rho(1-hz)}{1-bz},&
 \mu_S(z)&=\rho. \label{eq:historypost}
\end{align}
In particular,
\[
 \mu_-(z)<\rho<\mu_+(z).
\]
Conditional on risky advice, average project quality is
\begin{equation}
 \bar p_R(z)=\frac12(1+bz),\qquad
 \bar p_R(z)-p(z)=\frac{b(1-z)}2>0. \label{eq:selectiongap}
\end{equation}
Equation~\eqref{eq:selectiongap} is the selection force behind the main
contracting result.

\section{Career concerns and laissez faire}\label{sec:reputation}

Before any contract is written, it is worth asking whether the distortion the
paper proposes to fix is really there. It is, and it comes from a single
feature of the environment: safe advice generates no evidence about ability,
and no evidence is worth the prior.\footnote{This is the point at which our
wedge parts company with a mechanical ``failure hurts more than success
helps'' intuition. Reputation is a martingale, so the two outcome posteriors
straddle the prior and are correctly weighted by the marginal adviser's own
beliefs. What breaks the tie is that those weights are the \emph{marginal}
adviser's, while the posteriors are generated by the whole selected population
of risky recommendations---the marginal adviser is more pessimistic than the
average one, and therefore expects to lose reputation on net.}

At candidate confidence cutoff \(c\), define the pure career wedge
\begin{equation}
 K(c)=\gamma\{\rho-c\mu_+(c)-(1-c)\mu_-(c)\}. \label{eq:Kdef}
\end{equation}
This is the career value of safe advice minus the expected career value of
risky advice for the marginal adviser. Substitution from
\eqref{eq:historypost} gives
\begin{equation}
 K(z)=
 \gamma
 \frac{\rho(1-\rho)(h-l)b\,z(1-z)}
 {1-b^2z^2}>0. \label{eq:Kclosed}
\end{equation}
The wedge is strictly concave in both \(z\) and \(c\). It vanishes only
at the boundaries, where the recommendation ceases to select
ability.\footnote{The two boundary cases are instructive. If the adviser
always recommends risk, the recommendation carries no information about her
signal and, under the boundary limits used below, all three histories return
the prior; if she never does, the outcome histories are never reached. In
both cases there is nothing for the market to learn from the recommendation,
and the career motive is silent. Reputational conservatism is thus a strictly
interior phenomenon.}

The adviser's material gain from risky advice at confidence \(c\) is
\(cg-(1-c)\ell\). Define the total encouragement deficit and required
failure protection:
\begin{align}
 D(c)&=K(c)-\{cg-(1-c)\ell\}, \label{eq:Ddef}\\
 \sigma(c)&=\frac{D(c)}{1-c}. \label{eq:sigma}
\end{align}
By proportional alignment,
\[
 cg-(1-c)\ell=(g+\ell)(c-\cFB).
\]
The laissez-faire cutoff solves \(D(c)=0\).

\begin{proposition}\label{prop:lf}
There is a unique laissez-faire cutoff
\[
 \cLF\in(\cFB,\obar).
\]
Thus career concerns suppress some positive-value risky recommendations.
\end{proposition}

The intuition is immediate at first best. Material incentives are exactly
balanced there---this is what proportional alignment buys us---but
\(K(\cFB)>0\), so the marginal adviser prefers safety.
At the top of the support, the career wedge vanishes and the aligned
material gain from risk is positive. Strict concavity gives a unique
crossing to the right of first best.

To implement targets globally over nonpooling strategies, we impose a
primitive condition ensuring that success remains preferable to safety
even under the most unfavorable selection-induced reputation:
\begin{assumption}[Uniform success dominance]\label{ass:usd}
\begin{equation}
 g>
 \gamma\frac{\rho(1-\rho)(h-l)}{1-b}. \tag{USD}\label{eq:usd}
\end{equation}
\end{assumption}
Under Assumption~\ref{ass:usd}, \(\sigma'(c)<0\) throughout the confidence
support. A more cautious target requires less downside
protection.\footnote{Two things should be said about \eqref{eq:usd}. It is a
joint restriction on the size of the material stake relative to the career
stake, and it is what rules out the perverse best response in which an adviser
recommends risk precisely when her confidence is \emph{low}; it is not a
restriction on the sign of the wedge, which Proposition~\ref{prop:lf} obtains
without it. Note also that the monotonicity of \(\sigma\) is doing double
duty in what follows: it delivers uniqueness of the implemented cutoff, and it
is the reason the marginal cost of intervention can be evaluated at the
observed laissez-faire policy in Section~\ref{sec:principal}.}

\section{Least-cost incentives and implementation}\label{sec:contract}

The contracting problem can now be stated, and it is deliberately stated
first in a form that does not depend on the canonical density: the cost
comparison at the heart of the paper is a selection argument, and selection
arguments do not need functional forms. Let posterior confidence \(p\) have any atomless,
nondegenerate distribution. An upper-cutoff rule recommends risk when
\(p\geq c\). Let \(P_+(c)\), \(P_-(c)\), and \(P_R(c)\) be the
probabilities of success, failure, and risky advice.

At the marginal confidence, indifference requires
\begin{equation}
 ct_+ +(1-c)t_- -t_S=D(c). \label{eq:targetIC}
\end{equation}

\begin{theorem}\label{thm:selection}
Fix an interior upper cutoff \(c\) with \(D(c)>0\). Among all nonnegative
incremental payments that support the cutoff under the beliefs it
generates, the unique least-cost contract is
\begin{equation}
 t_S^*(c)=t_+^*(c)=0,\qquad
 t_-^*(c)=\frac{D(c)}{1-c}. \label{eq:leastcost}
\end{equation}
It is strictly cheaper than a success-only contract. The expected-cost
difference is
\begin{equation}
 C_+(c)-C_-(c)
 =
 D(c)P_R(c)
 \frac{\E[p\mid p\geq c]-c}{c(1-c)}
 >0. \label{eq:costgap}
\end{equation}
\end{theorem}

To see the logic, one unit of marginal incentive delivered through success
pay costs
\[
 \frac{P_+(c)}{c}
 =
 P_R(c)\frac{\E[p\mid p\geq c]}{c},
\]
whereas one unit delivered through failure protection costs
\[
 \frac{P_-(c)}{1-c}
 =
 P_R(c)\frac{1-\E[p\mid p\geq c]}{1-c}.
\]
Upper-tail selection makes the latter strictly smaller. A safe payment is
both costly and counterproductive in \eqref{eq:targetIC}: it has to be paid,
and it makes safety more attractive, which is the opposite of what the
organization wants. The result is therefore a linear-programming consequence
of selection, not a claim that the adviser is intrinsically averse to
risk.\footnote{Read the two expressions as prices. The organization is buying
one unit of incentive at the margin, and it may pay for it in either of two
currencies; the exchange rate is set by how often each currency has to be
handed over across the whole selected population. Because that population is
better than marginal, successes are relatively common and failures relatively
rare, which makes failure the cheap currency. The size of the discount,
recorded in \eqref{eq:costgap}, is exactly the selection gap
\(\E[p\mid p\geq c]-c\).}

The next result closes the equilibrium problem.

\begin{theorem}\label{thm:implementation}
Suppose \eqref{eq:interior} and Assumption~\ref{ass:usd} hold. For every
\(c\in(\ubar,\obar)\) with \(D(c)>0\), the contract in
\eqref{eq:leastcost} induces the target upper cutoff as the unique
nonpooling equilibrium among all measurable, possibly mixed advice
strategies.
\end{theorem}

The proof has two steps. First, the risky--safe payoff difference is affine
in posterior confidence. Uniform success dominance rules out a decreasing
or flat best response under any nonpooling selection, so every nonpooling
equilibrium is an upper cutoff. Second, an alternative cutoff \(q\) must
satisfy
\[
 \sigma(q)=\sigma(c).
\]
The strict decrease of \(\sigma\) implies \(q=c\).

Pooling requires a separate qualification, and we would rather state it than
finesse it, because outcome beliefs are off path at all-safe advice. Under \emph{upper-tail-consistent boundary
beliefs}---the continuous limits generated by upper cutoffs approaching
the two boundaries---both all-safe and all-risk are unstable. The target
is then the unique equilibrium, up to behavior at the null cutoff signal.
Without that explicit refinement, Theorem~\ref{thm:implementation} is the
refinement-free result.

\section{A cap on downside protection}\label{sec:cap}

The least-cost contract of Section~\ref{sec:contract} asks an organization to
do the one thing organizations find hardest to justify, which is to pay after
a visible failure. It is therefore worth asking what happens when that
instrument is available but bounded. Suppose failure protection cannot exceed
\(\bar t\geq0\), while success pay remains uncapped. Such a constraint may
capture limited commitment, internal-pay equity, clawback rules, or the
reputational difficulty of explaining payments made after a failed
project.\footnote{We treat the cap as exogenous, which is the honest way to
model a governance constraint whose origin lies outside the interaction we
study. The relevant comparative static is nevertheless the one an
organization can act on: \(\bar t\) indexes how credibly downside protection
can be promised, and Proposition~\ref{prop:cap} together with
\eqref{eq:capcost} says what is lost at each level of credibility.}

\begin{proposition}\label{prop:cap}
Fix an interior target \(c\) with \(D(c)>0\). The unique least-cost
incremental contract subject to \(t_-\leq\bar t\) is
\begin{align}
 t_S^*(c;\bar t)&=0,\\
 t_-^*(c;\bar t)&=\min\{\bar t,\sigma(c)\},\\
 t_+^*(c;\bar t)
 &=\frac{1-c}{c}[\sigma(c)-\bar t]_+ . \label{eq:capcontract}
\end{align}
It uniquely implements the target among nonpooling equilibria and fully
implements it under upper-tail-consistent boundary beliefs.
\end{proposition}

Failure protection is exhausted before success pay is used---the ordering,
not merely the mix, is the content of the proposition. Let
\[
 C_\infty(c)=P_-(c)\sigma(c)
\]
be expected compensation without a cap and define
\[
 Q(c)=
 \frac{1-c}{c}P_+(c)-P_-(c)
 =
 P_R(c)\frac{\bar p_R(c)-c}{c}>0.
\]
Expected compensation under the cap is
\begin{equation}
 C_{\bar t}(c)
 =
 C_\infty(c)+Q(c)[\sigma(c)-\bar t]_+. \label{eq:capcost}
\end{equation}
It falls strictly as the cap is relaxed while the cap binds and is
constant thereafter. On the encouragement region
\([\cFB,\cLF)\), compensation is strictly decreasing in \(c\). The
smallest and largest optimal cutoffs are therefore weakly decreasing in
\(\bar t\): an organization that can credibly provide more downside
protection induces weakly bolder advice.

\section{The organization's target}\label{sec:principal}

So far the target cutoff has been fixed and the question has been how cheaply
to reach it. We now let the organization choose the target too, which is where
the design problem acquires a second margin and where the least-cost result
stops being the end of the story. Gross project surplus under confidence
cutoff \(c\) is
\begin{equation}
 B(c)=
 \int_c^{\obar}\{pG-(1-p)L\}\,dF(p), \label{eq:benefit}
\end{equation}
and
\[
 B'(c)=-(G+L)(c-\cFB)f(c).
\]
The organization chooses
\[
 \max_c W_{\bar t}(c)
 =
 \max_c\{B(c)-C_{\bar t}(c)\}.
\]

\begin{theorem}\label{thm:location}
An optimal target exists, and every optimal target satisfies
\begin{equation}
 c^*\in(\cFB,\cLF]. \label{eq:location}
\end{equation}
The organization either leaves laissez faire unchanged or corrects the
career distortion only partially. Exact first-best correction is never
optimal.
\end{theorem}

A cutoff below first best implements negative-value projects and costs
more than first best. A cutoff above laissez faire excludes
positive-value projects and cannot cost less than the zero-payment
laissez-faire outcome. Finally,
\[
 B'(\cFB)=0
 \quad\text{while}\quad
 C_{\bar t}'(\cFB+)<0.
\]
Starting at first best, a marginal increase in caution saves compensation
to first order but loses project surplus only to second order. The
organization therefore chooses \(c>\cFB\).\footnote{The asymmetry of orders is
worth pausing on, because it is the reason the conclusion is a
\emph{never}\/ rather than a \emph{sometimes}. At \(\cFB\) the marginal
project is exactly break-even by construction, so the surplus consequence of
excluding it is nil at first order; the compensation consequence is not, since
the organization is still paying \(\sigma(\cFB)>0\) on every failure. The
same local logic applies whenever efficiency is defined by a break-even margin
and correction remains costly at that margin, so we would not read partial
correction as a delicate result.}

Theorem~\ref{thm:location} locates the optimum but does not pin it down, and
uniqueness needs curvature. We would rather obtain that curvature from
primitives than assume it, and in a natural version of the baseline we can.
The condition \(\cFB\geq1/2\) is equivalent to \(L\geq G\): failure is at
least as costly to the organization as success is valuable.

\begin{proposition}
\label{prop:primitivecurvature}
Suppose the canonical signal model is uncapped, Assumption~\ref{ass:usd}
holds, and \(\cFB\geq1/2\). Then \(C_\infty\) is strictly convex on
\([\cFB,\cLF]\), and the organization's objective \(W_\infty\) is
strictly concave there.
\end{proposition}

Thus uniqueness in the canonical uncapped problem does not require a
separate curvature assumption. The Appendix gives a weaker
one-dimensional primitive sufficient condition; \(\cFB\geq1/2\) is a
convenient transparent specialization, not a necessary
condition.\footnote{The restriction has a transparent payoff interpretation:
failure is at least as costly to the organization as success is valuable. The
weaker primitive sufficient condition stated in the Appendix can also
hold with \(L<G\), so none of the design conclusions turns on loss dominance.}

To state the baseline and its extensions in a common form, define the
marginal project loss
\[
 \Psi(c)=(G+L)(c-\cFB)f(c)
\]
and the marginal compensation saving
\[
 M_{\bar t}(c)=-C_{\bar t}'(c).
\]

Proposition~\ref{prop:primitivecurvature} makes \(\Psi\) strictly
increasing and \(M_\infty\) strictly decreasing in the uncapped
canonical baseline. For a binding cap or a noncanonical confidence
distribution, we use the transparent \emph{marginal single-crossing}
condition that \(\Psi\) is strictly increasing and \(M_{\bar t}\) is
weakly decreasing on \([\cFB,\cLF]\).

\begin{theorem}\label{thm:optimum}
Suppose either that the primitive baseline conditions in
Proposition~\ref{prop:primitivecurvature} hold or that marginal single
crossing holds. The optimal target is unique. The organization
intervenes if and only if
\begin{equation}
 (G+L)(\cLF-\cFB)f(\cLF)>M_{\bar t}(\cLF-).
 \label{eq:inttest}
\end{equation}
If the inequality holds, \(c^*\in(\cFB,\cLF)\) is the unique solution to
\[
 (G+L)(c^*-\cFB)f(c^*)=M_{\bar t}(c^*).
\]
Otherwise \(c^*=\cLF\).
\end{theorem}

The endpoint test is useful for a practical reason: the cost of a small
intervention can be evaluated at the observed laissez-faire policy, which is
the one policy an outside observer can hope to measure. For every positive
finite cap, as well as in the uncapped problem, failure protection is
locally unconstrained near \(\cLF\), and
\[
 M_{\bar t}(\cLF-)=-P_-(\cLF)\sigma'(\cLF).
\]
If failure protection is prohibited, the corresponding marginal cost is
\[
 -\frac{1-\cLF}{\cLF}P_+(\cLF)\sigma'(\cLF),
\]
which is strictly larger by upper-tail selection. Restricting the
organization to conventional success pay can therefore turn a profitable
intervention into laissez faire.\footnote{This is the sharpest observable
implication of the selection logic, and it is a comparison between
institutions rather than between contracts: two organizations with identical
technologies, identical advisers, and identical career exposure can display
different advice thresholds purely because one of them is permitted to protect
failure and the other is not. Excessive caution need not indicate weak
governance; it can indicate governance that has ruled out the cheap
instrument.}

\section{Internal review and disclosure}\label{sec:review}

Compensation is not the only instrument an organization has, and it is
usually not the first one it reaches for. Recommendations are screened before
they are implemented, and how much of that screening the outside world gets to
see is itself a choice. We now allow an internal reviewer to screen a risky
recommendation before implementation. A bad project is detected and vetoed with probability
\(\lambda\in[0,1]\); a good project is never vetoed. Conditional on a bad
project and risky advice, detection is independent of ability and of the
precise signal. A veto produces zero material payoff.

Under \emph{transparent review}, the outside market observes
\(\{S,V,+,-\}\), including a veto \(V\). Under \emph{confidential review},
the market observes \(\{N,+,-\}\), where \(N\) pools safe advice and an
internal veto. The employer observes the underlying event, can contract
on it, and keeps compensation private.

Fix raw-signal cutoff \(z\), and write \(q=p(z)\). A veto and an
implemented failure have proportional type likelihoods---both occur exactly
when a bad project was recommended, and detection is independent of
ability---so transparent review gives
\[
 \mu_V(z,\lambda)=\mu_-(z).
\]
Under confidential review, let
\[
 m_\theta(z,\lambda)
 =
 z+\frac{\lambda}{2}(1-z)(1-a_\theta z).
\]
The posterior following no launch is
\begin{equation}
 \mu_N(z,\lambda)=
 \frac{\rho m_H(z,\lambda)}
 {\rho m_H(z,\lambda)+(1-\rho)m_L(z,\lambda)}. \label{eq:muN}
\end{equation}
For \(\lambda>0\),
\[
 \mu_-(z)<\mu_N(z,\lambda)<\rho,
\qquad
 \frac{\partial\mu_N}{\partial\lambda}<0.
\]

Let \(K_T(z,\lambda)\) and \(K_C(z,\lambda)\) denote the marginal career
wedges under transparent and confidential review. With
\(v_h=V(\mu_h)\),
\begin{align}
 K_T(z,\lambda)
 &=
 v_S-\{qv_++(1-q)v_-\}, \label{eq:KT}\\
 K_C(z,\lambda)
 &=
 v_N-
 \left[qv_+
 +(1-q)\{(1-\lambda)v_-+\lambda v_N\}\right]. \label{eq:KC}
\end{align}

\begin{proposition}\label{prop:shield}
Transparent review leaves the pure career wedge unchanged:
\[
 \frac{\partial K_T}{\partial\lambda}=0.
\]
Confidential review strictly reduces it:
\[
 \frac{\partial K_C}{\partial\lambda}<0.
\]
Hence \(K_C(z,\lambda)<K_T(z,\lambda)\) for every \(\lambda>0\), and
the gap grows with review precision.
\end{proposition}

Review also prevents the adviser's material loss after some bad
recommendations. Under disclosure regime \(j\in\{T,C\}\), the total
encouragement deficit becomes
\[
 D_j(z,\lambda)=
 K_j(z,\lambda)-\{qg-(1-q)(1-\lambda)\ell\}.
\]
Both regimes reduce the deficit, but confidentiality reduces it faster.
Let \(t_V\geq0\) denote a privately observed payment following a veto.

\begin{theorem}\label{thm:review}
Fix \((z,\lambda)\) and suppose \(D_j(z,\lambda)>0\). Every least-cost
schedule sets \(t_S=t_+=0\) and supplies the required incentive through
failure and veto protection:
\[
 (1-\lambda)t_-+\lambda t_V
 =
 \frac{D_j(z,\lambda)}{1-q}.
\]
All such schedules have expected cost
\begin{equation}
 C_j^I(z,\lambda)
 =
 (1-z)\frac{1-\bar p_R(z)}{1-q}D_j(z,\lambda).
 \label{eq:reviewcost}
\end{equation}
While \(D_C>0\), confidential review is strictly cheaper than transparent
review, and its cost advantage increases with \(\lambda\).
\end{theorem}

\begin{corollary}\label{cor:reviewintensity}
Fix \(z\) and a compact set of feasible review precisions on which
\(D_C(z,\lambda)>0\). Suppose review has the same real cost and generates
the same project surplus under the two disclosure regimes. Then every
optimal review precision under confidentiality is weakly greater than
every optimal review precision under transparency.
\end{corollary}

The corollary does not require a particular review-cost function. On the
stated domain, the organization's payoff under confidentiality equals its
payoff under transparency plus a compensation-cost advantage that is
strictly increasing in review precision. Transparency therefore cannot
select a strictly higher review precision than confidentiality.

The domain restriction matters, and we would rather advertise it than bury
it. If review is strong enough to make the deficit negative, the fixed target
requires a safe payment to restrain risky advice; because confidentiality
reduces the deficit further, it can then be more expensive. We therefore do
not claim that secrecy dominates transparency globally, and the numerical
illustration in Section~\ref{sec:numerical} reports where the boundary lies in
the benchmark rather than leaving the reader to guess that it is
far away.\footnote{It is not far away. At the second-best target of the
benchmark the confidential-review deficit vanishes at a detection probability
of roughly five percent, and the transparent-review deficit at roughly seven
percent, so the region in which the ranking is claimed is genuinely small
there. We regard this as a reason to state the result as a local
substitution property rather than as a disclosure policy.} The result is instead an organizational
substitution theorem: while career concerns are suppressing desirable
risky advice, confidential review supplies career insurance that
transparent review does not.

\section{Numerical illustration}\label{sec:numerical}

The illustration below is not evidence for anything; it is a check that the
objects in the theorems have plausible magnitudes and that the two margins of
the design problem can be seen separately. Consider
\[
 h=.8,\quad l=.2,\quad \rho=\frac12,\quad\gamma=1,
 \quad g=\ell=.4,\quad G=L=10.
\]
The uniform success-dominance bound is \(0.3<g\). The efficient raw-signal
cutoff, laissez-faire cutoff, and optimal cutoff without a cap are
\[
 z^{\mathrm{FB}}=.500000,\qquad
 z^0=.550202,\qquad
 z^*=.515377.
\]
At the optimum,
\[
 p(z^*)=.507689,\qquad
 \bar p_R(z^*)=.628844,
\]
so the recommendations receiving payment are positively selected relative
to the margin. The required deficit and failure payment are
\[
 D(z^*)=.013914,\qquad t_-^*(z^*)=.028262.
\]
Expected failure-protection cost is \(0.005083\), compared with \(0.008352\)
under success-only pay---a reduction of approximately
\(39\%\).\footnote{The magnitude of the saving is entirely a statement about
selection, and it can be read off \eqref{eq:costgap}: at the optimum the
selection gap \(\bar p_R-p\) is about twelve percentage points on a base
confidence of about one half, and the cost ratio inherits it. A tighter
target---one closer to laissez faire---has a smaller deficit but the same
qualitative discount.}

\begin{figure}[t]
\centering
\includegraphics[width=.88\textwidth]{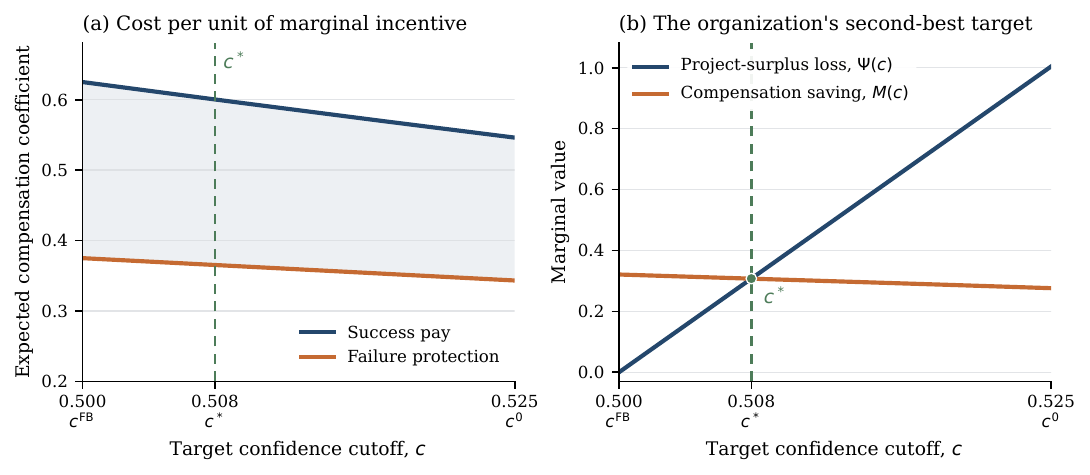}
\caption{Selection and the second-best target in the numerical benchmark.
Panel (a) compares expected compensation per unit of marginal incentive:
upper-tail selection makes failure protection cheaper. Panel (b) plots
the marginal project-surplus loss \(\Psi(c)\) and compensation saving
\(M_\infty(c)\); their intersection is \(c^*\), marked in green.}
\label{fig:selectiondesign}
\end{figure}

Figure~\ref{fig:selectiondesign} separates the paper's two margins. Panel
(a) holds the target fixed and compares instruments. Panel (b) lets the
organization move the target: at first best the project-surplus loss is
zero but compensation savings are positive, while at laissez faire the
project-surplus loss is larger. The crossing produces partial rather than
complete correction.

At this same target, the confidential-review deficit reaches zero near
\(\lambda=.0492\), while the transparent-review deficit reaches zero near
\(\lambda=.0707\). The positive-deficit comparison in
Theorem~\ref{thm:review} is therefore illustrated on
\(\lambda\in[0,.04]\). The replication code verifies the closed forms,
equilibrium inequalities, cost comparisons, and review-domain boundaries.

\FloatBarrier
\section{Conclusion}\label{sec:conclusion}

When advisers are evaluated through the outcomes of the projects they
recommend, even advisers who want exactly what the organization wants will
suppress worthwhile risk. The organization's cheapest response is not to pay
for success. Because risky recommendations select the upper tail of private
confidence, success pay is concentrated on recommendations that are already
inframarginal, and protecting the adviser following failure moves the advice
threshold at strictly lower expected cost. The central cost comparison is a
selection argument: it needs no risk aversion, no dynamics, and no private
knowledge of ability.

The design implications are simple enough to state as a rule of thumb: use
downside protection where it is available; add success pay only where
protection is capped, and in that order; and recognize that confidential
internal review can do part of the work of explicit career insurance. One
implication is worth emphasizing because it runs against instinct---the
optimal policy does not eliminate reputational caution. Exact first-best
implementation is too expensive at the margin, so intervention is either
absent or partial, and an organization that observes its advisers behaving
somewhat too cautiously has not thereby shown that eliminating all caution
would be worth its cost.

The mechanism also yields predictions that could in principle be told apart
from ordinary risk sharing, which is absent from the model. Holding a target
fixed, stronger career exposure raises the demand for downside protection. A
tighter cap shifts compensation toward success pay and advice toward greater
caution. Confidential review and explicit career insurance act as
substitutes, with confidentiality supporting greater review intensity
precisely while the organization is trying to encourage risk.\footnote{The
cleanest of these to look for is the cap comparative static, since it predicts
a joint movement---compensation toward success-contingent pay
\emph{and} advice toward caution---in response to something, such as a
clawback rule or a pay-equity constraint, that is often imposed for reasons
having nothing to do with project selection.}

Two limitations should be stated plainly rather than left to a referee.
First, the analysis deliberately separates advice from hidden effort and from
dynamic experimentation, and both extensions would change the object being
contracted on rather than merely complicate it. Second, and more
consequentially, ability is commonly unknown here. A natural next step is to
study privately known ability, where advisers select simultaneously on project
confidence and on self-assessment; that extension may create screening forces
absent here, and it may not preserve even the sign of the distortion, since
\citet{Chen2015} shows that private knowledge of ability can turn
underinvestment in risky projects into overinvestment. We would nevertheless
argue for evaluating it only after the clean common-uncertainty benchmark is
fully understood: if the least-cost instrument is already
counterintuitive when the reputational wedge has an unambiguous sign, there is
little to be gained from meeting it first in a setting where it does not.

\appendix

\section{Environment}

We restate the environment in full, so that the appendix can be read on its
own. Notation follows the manuscript throughout, with one exception noted
where it arises: to keep the confidence distribution \(F\) free, the curvature
function used in Proposition~\ref{prop:primitivecurvature} is written
\(\Phi\), and the aggregate stake \(g+\ell\) is written \(s\).

Ability and project quality are binary:
\[
 \theta\in\{H,L\},\qquad \Prb(\theta=H)=\rho\in(0,1),
 \qquad
 \omega\in\{1,0\},\qquad \Prb(\omega=1)=\frac12.
\]
They are independent, and ability is initially unknown to both the
adviser and the market. Conditional on $(\theta,\omega)$, the adviser
observes $x\in[0,1]$ with density
\begin{align}
 f_\theta(x\mid 1)&=1+a_\theta(2x-1),&
 f_\theta(x\mid 0)&=1-a_\theta(2x-1), \label{eq:density}
\end{align}
where
\[
 0\leq l\equiv a_L<h\equiv a_H<1,\qquad
 b\equiv \rho h+(1-\rho)l\in(0,1).
\]

The adviser recommends a safe action $S$ or a risky action $R$. Risky
advice is implemented. A successful risky project generates $G>0$ for
the organization and a failed one generates $-L<0$; safety yields zero.
The organization's efficient posterior cutoff is
\[
 \cFB=\frac{L}{G+L}.
\]
The adviser obtains $g>0$ after success, $-\ell<0$ after failure, and
zero after safety. We impose proportional alignment:
\begin{equation}
 \frac{\ell}{g+\ell}=\frac{L}{G+L}=\cFB. \label{eq:alignment}
\end{equation}
Thus the adviser and the organization agree about which projects are
efficient absent career concerns, although their payoff levels may
differ.

The market observes $S$, a successful risky recommendation $+$, or a
failed risky recommendation $-$. If $\mu_h$ is the posterior
probability of high ability after history $h$, the adviser's career
value is
\[
 V(\mu_h)=\gamma\mu_h,\qquad \gamma>0.
\]
The organization may offer nonnegative incremental payments
$(t_S,t_+,t_-)\geq0$. They are paid above an immutable base salary; the
adviser is an incumbent and participation is slack for the incremental
contract.

\begin{assumption}\label{ass:interior}
The efficient cutoff lies in the interior of the posterior support:
\[
 \frac{1-b}{2}<\cFB<\frac{1+b}{2}.
\]
\end{assumption}

\begin{assumption}[Uniform success dominance]\label{ass:usd}
\begin{equation}
 g>
 \gamma\frac{\rho(1-\rho)(h-l)}{1-b}. \tag{USD}\label{eq:usd}
\end{equation}
\end{assumption}

Assumption~\ref{ass:usd} says that an actual success remains better for
the adviser than safe advice even when the selection of risky
recommendations makes success as unfavorable a reputation signal as the
technology permits. The bound on the right-hand side is attained in the limit
and is therefore not conservative by more than the technology
allows.\footnote{Lemma~\ref{lem:arbitrary} shows that
\(\rho-\mu_+(r)\) cannot exceed \(\rho(1-\rho)(h-l)/(1-b)\) under any
measurable advice rule, and that this infimum of \(\mu_+\) is approached as the
rule concentrates on low signals. Assumption~\ref{ass:usd} is thus the
weakest uniform condition of its kind: any strictly smaller \(g\) admits a
selection under which success is reputationally worse than safety, and the
cutoff characterization of best responses then fails.}

\section{Information and reputational updating}

The first two lemmas are bookkeeping, but they are the bookkeeping the rest of
the appendix relies on: Lemma~\ref{lem:signal} establishes that higher ability
is more information in the Blackwell sense and that posterior confidence is
uniform, and Lemma~\ref{lem:histories} records the three history
probabilities, the three reputations, and the selection gap that drives the
cost comparison.

\begin{lemma}\label{lem:signal}
The densities in \eqref{eq:density} are strictly positive and integrate
to one. The signal satisfies the monotone likelihood-ratio property.
The low-ability experiment is a Blackwell garbling of the high-ability
experiment. Posterior project quality is
\begin{equation}
 p(x)\equiv\Prb(\omega=1\mid x)
 =\frac{1-b}{2}+bx. \label{eq:posterior}
\end{equation}
Consequently, $p$ is strictly increasing and uniformly distributed on
\[
 [\ubar,\obar]
 \equiv
 \left[\frac{1-b}{2},\frac{1+b}{2}\right].
\]
\end{lemma}

\begin{proof}
Positivity and unit mass follow directly from \eqref{eq:density}. For
$a>0$, the state likelihood ratio
\[
 \frac{1+a(2x-1)}{1-a(2x-1)}
\]
has derivative
\[
 \frac{4a}{[1-a(2x-1)]^2}>0;
\]
for $a=0$ it is constant. To obtain the low-ability experiment from the
high-ability experiment, retain a high-ability signal with probability
$l/h$ and otherwise replace it by an independent uniform draw. The
resulting density is
\[
 \frac{l}{h}f_H(x\mid\omega)+\left(1-\frac{l}{h}\right)
 =f_L(x\mid\omega).
\]
After integrating over ability, the state-contingent densities are
$1+b(2x-1)$ and $1-b(2x-1)$. Bayes' rule with an equal state prior gives
\eqref{eq:posterior}. Finally, averaging \eqref{eq:density} over the two
states gives the uniform density for either ability, and hence also
unconditionally. The affine transformation in \eqref{eq:posterior}
therefore makes $p$ uniform on the stated support.
\end{proof}

Write $z\in(0,1)$ for a raw-signal cutoff and
\[
 c=p(z)=\ubar+bz
\]
for its posterior counterpart. Under the upper-cutoff rule
$R\Longleftrightarrow x\geq z$, define $P_h(z)$ as the ex ante
probability of history $h$.

\begin{lemma}\label{lem:histories}
For an upper cutoff $z$,
\begin{align}
 P_+(z)&=\frac12(1-z)(1+bz),&
 P_-(z)&=\frac12(1-z)(1-bz),&
 P_S(z)&=z, \label{eq:historyprob}\\
 \mu_+(z)&=\frac{\rho(1+hz)}{1+bz},&
 \mu_-(z)&=\frac{\rho(1-hz)}{1-bz},&
 \mu_S(z)&=\rho. \label{eq:historypost}
\end{align}
For every interior cutoff,
\[
 \mu_-(z)<\rho<\mu_+(z).
\]
Moreover,
\begin{align}
 \bar p_R(z)
 &\equiv\Prb(\omega=1\mid x\geq z)
 =\frac12(1+bz), \label{eq:avgp}\\
 \bar p_R(z)-p(z)&=\frac{b(1-z)}{2}>0. \label{eq:selectiongap}
\end{align}
\end{lemma}

\begin{proof}
Conditional on type $\theta$, the probabilities of risky advice in the
two states are
\[
 r_\theta^1(z)=(1-z)(1+a_\theta z),\qquad
 r_\theta^0(z)=(1-z)(1-a_\theta z).
\]
Averaging over ability and multiplying by the equal state probabilities
gives \eqref{eq:historyprob}. Bayes' rule gives
\eqref{eq:historypost}. Since $h>b$ and $b>\!l$, the stated posterior
ranking follows. Dividing $P_+$ by $P_++P_-$ proves
\eqref{eq:avgp}; subtracting $p(z)$ proves
\eqref{eq:selectiongap}.\footnote{The uniformity of $p$ is a convenience of
the linear family and is used only in the canonical results of Sections~7
and~8. Theorem~\ref{thm:selection} requires only that the confidence
distribution be atomless and nondegenerate above the cutoff. The unit-cost
comparison in Theorem~\ref{thm:reviewcost} uses the same selection property,
while its disclosure ranking also uses the review structure introduced
later.}
\end{proof}

\section{The reputational wedge and laissez faire}

The wedge is strictly positive at every interior cutoff and vanishes at both
boundaries, where the recommendation stops carrying information about ability.
Its strict concavity, established below, is used twice: it makes the
laissez-faire cutoff unique, and, through Lemma~\ref{lem:monosigma}, it makes
the required failure protection strictly decreasing in the target.

For a candidate posterior cutoff $c$, define
\begin{align}
 K(c)
 &\equiv
 \gamma\{\rho-c\mu_+(c)-(1-c)\mu_-(c)\}, \label{eq:Kdef}\\
 D(c)
 &\equiv K(c)-\{cg-(1-c)\ell\}, \label{eq:Ddef}\\
 \sigma(c)
 &\equiv \frac{D(c)}{1-c}. \label{eq:sigmadef}
\end{align}
Here and below $\mu_h(c)$ means \eqref{eq:historypost} evaluated at
$z=(c-\ubar)/b$. The term $K$ is the pure career-concern wedge, $D$ is
the total encouragement deficit, and $\sigma$ is the failure protection
required at the margin.

\begin{proposition}\label{prop:wedge}
For $z\in(0,1)$,
\begin{equation}
 K(z)=
 \gamma
 \frac{\rho(1-\rho)(h-l)b\,z(1-z)}
 {1-b^2z^2}>0. \label{eq:Kclosed}
\end{equation}
The function $K$ is strictly concave in both $z$ and $c$.
\end{proposition}

\begin{proof}
Substituting \eqref{eq:historypost} and $c=\ubar+bz$ into
\eqref{eq:Kdef} gives \eqref{eq:Kclosed}. Apart from a positive
constant, its variable part is
\[
 k(z)=\frac{z(1-z)}{1-b^2z^2}.
\]
Direct differentiation gives
\[
 k''(z)=
 -\frac{2[1-3b^2z+3b^2z^2-b^4z^3]}
 {(1-b^2z^2)^3}.
\]
The bracket is strictly positive because
\[
 1-3b^2z+3b^2z^2-b^4z^3
 =(1-b^2z)^3+b^2(1-b^2)z^2(3-b^2z)>0.
\]
Thus $K$ is strictly concave in $z$. Since $c$ is a positive affine
transformation of $z$, it is strictly concave in $c$ as well.
\end{proof}

\begin{proposition}\label{prop:lf}
Under Assumption~\ref{ass:interior}, the equation $D(c)=0$ has exactly
one solution $\cLF$ in $(\cFB,\obar)$. The laissez-faire upper cutoff is
therefore reputationally conservative:
\[
 \cLF>\cFB.
\]
\end{proposition}

\begin{proof}
Proportional alignment implies
\[
 cg-(1-c)\ell=(g+\ell)(c-\cFB).
\]
Hence $D(c)>0$ for $c\leq\cFB$, with
$D(\cFB)=K(\cFB)>0$. At the upper endpoint, $K(\obar)=0$ and
\[
 D(\obar)=-(g+\ell)(\obar-\cFB)<0.
\]
By Proposition~\ref{prop:wedge}, $D$ is strictly concave. Its
nonnegative upper contour set is therefore an interval; because it
contains $[\ubar,\cFB]$ but not $\obar$, it has a unique right endpoint
$\cLF\in(\cFB,\obar)$. This endpoint is the unique zero to the right of
first best.
\end{proof}

\begin{lemma}\label{lem:monosigma}
Under Assumption~\ref{ass:usd},
\[
 \sigma'(c)<0\qquad\text{for all }c\in[\ubar,\obar].
\]
\end{lemma}

\begin{proof}
Differentiating \eqref{eq:sigmadef} and using proportional alignment
gives
\begin{equation}
 \sigma'(c)
 =
 \frac{K(c)+(1-c)K'(c)-g}{(1-c)^2}. \label{eq:sigmaprime}
\end{equation}
Because $K''<0$, the function
$K(c)+(1-c)K'(c)$ is strictly decreasing. At the lower endpoint,
\[
 K(\ubar)=0,\qquad
 K'(\ubar)=\gamma\rho(1-\rho)(h-l),
\]
so its maximum is
\[
 \gamma
 \frac{\rho(1-\rho)(h-l)(1+b)}{2}.
\]
Since $1/(1-b)>(1+b)/2$, Assumption~\ref{ass:usd} makes the numerator in
\eqref{eq:sigmaprime} strictly negative throughout the
support.\footnote{Assumption~\ref{ass:usd} is deliberately stronger than the
condition needed for this lemma. The weaker bound
\(g>\gamma\rho(1-\rho)(h-l)(1+b)/2\) suffices for
\(\sigma'<0\); Assumption~\ref{ass:usd} is imposed because it also rules out
lower-cutoff best responses under arbitrary selection in
Lemma~\ref{lem:endomon}.}
\end{proof}
\section{Least-cost support}

The central cost result is stated for an arbitrary confidence distribution,
because that is the level of generality at which its logic operates: it uses
only upper-tail selection, and therefore applies well beyond the
linear-density family of Section~2. Let $F$ be any atomless,
nondegenerate distribution of posterior confidence $p$, and let
\[
 P_+(c)=\Prb(p\geq c,\omega=1),\quad
 P_-(c)=\Prb(p\geq c,\omega=0),\quad
 P_R(c)=P_+(c)+P_-(c).
\]

\begin{theorem}\label{thm:selection}
Fix an interior upper cutoff $c$ with $D(c)>0$. Among all nonnegative
incremental payments that support the cutoff under the beliefs it
generates, the unique least-cost contract is
\begin{equation}
 t_S^*(c)=t_+^*(c)=0,\qquad
 t_-^*(c)=\frac{D(c)}{1-c}. \label{eq:leastcost}
\end{equation}
It is strictly cheaper than a success-only contract. In particular,
\begin{equation}
 C_+(c)-C_-(c)
 =
 D(c)P_R(c)
 \frac{\E[p\mid p\geq c]-c}{c(1-c)}
 >0. \label{eq:costgap}
\end{equation}
\end{theorem}

\begin{proof}
At the marginal confidence, indifference requires
\begin{equation}
 ct_+ +(1-c)t_- -t_S=D(c). \label{eq:targetIC}
\end{equation}
A positive safe payment both raises cost and relaxes
\eqref{eq:targetIC} in the wrong direction, so $t_S=0$ in any
least-cost schedule. One unit of marginal incentive delivered through
success pay costs
\[
 \frac{P_+(c)}{c}
 =
 P_R(c)\frac{\E[p\mid p\geq c]}{c},
\]
whereas one unit delivered through failure protection costs
\[
 \frac{P_-(c)}{1-c}
 =
 P_R(c)\frac{1-\E[p\mid p\geq c]}{1-c}.
\]
Because $F$ is atomless and nondegenerate above an interior cutoff,
\[
 \E[p\mid p\geq c]>c.
\]
Failure protection is therefore strictly cheaper per unit of marginal
incentive. The linear program uniquely allocates all required incentive
to $t_-$, which gives \eqref{eq:leastcost}. Comparing this contract with
$t_+=D(c)/c$ gives \eqref{eq:costgap}.
\end{proof}

\begin{remark}
The strict comparison concerns incremental performance pay. If a
freely adjustable base wage can be reduced and participation binds,
different outcome schedules may generate the same total wage bill.
Failure protection remains weakly cost minimizing, but strict total-cost
dominance need not survive.
\end{remark}

\section{Implementation over unrestricted advice strategies}

Theorem~\ref{thm:selection} supports a cutoff under the beliefs that cutoff
itself generates. That is not yet implementation, and this section closes the
gap by allowing the adviser any measurable, possibly mixed rule and asking
which such rules can survive.

An unrestricted behavioral advice strategy is a measurable function
$r:[0,1]\to[0,1]$, where $r(x)$ is the probability of risky advice. Its first
two signal moments are all that reputations depend on:
\[
 m_0(r)=\int_0^1r(x)\,dx,\qquad
 m_1(r)=\int_0^1r(x)(2x-1)\,dx.
\]
A strategy is nonpooling when $m_0(r)\in(0,1)$.

\begin{lemma}\label{lem:arbitrary}
For every nonpooling measurable strategy,
\begin{align}
 \mu_S(r)&=\rho, \label{eq:musany}\\
 \mu_+(r)
 &=\frac{\rho[m_0(r)+hm_1(r)]}{m_0(r)+bm_1(r)},&
 \mu_-(r)
 &=\frac{\rho[m_0(r)-hm_1(r)]}{m_0(r)-bm_1(r)}. \label{eq:muany}
\end{align}
Furthermore,
\begin{equation}
 \rho-\mu_+(r)
 \leq
 \frac{\rho(1-\rho)(h-l)}{1-b}. \label{eq:globalbound}
\end{equation}
\end{lemma}

\begin{proof}
Under the equal state prior, the unconditional density of $x$ is one
for either ability. Consequently, the probabilities of $S$ conditional
on $H$ and $L$ coincide, proving \eqref{eq:musany}. Conditional on type
$\theta$,
\begin{align*}
 \Prb(+\mid\theta)
 &=\frac12[m_0(r)+a_\theta m_1(r)],&
 \Prb(-\mid\theta)
 &=\frac12[m_0(r)-a_\theta m_1(r)].
\end{align*}
Bayes' rule gives \eqref{eq:muany}. Put
$\eta=m_1(r)/m_0(r)\in(-1,1)$. Then
\[
 \mu_+(r)=\rho\frac{1+h\eta}{1+b\eta},
\]
which is strictly increasing in $\eta$. Its infimum is
$\rho(1-h)/(1-b)$. Subtracting this value from $\rho$ yields
\eqref{eq:globalbound}.
\end{proof}

\begin{lemma}\label{lem:endomon}
Under Assumption~\ref{ass:usd}, every nonpooling equilibrium induced by
a failure-protection contract $(0,0,\tau)$ is, up to behavior at one
null signal, a deterministic upper cutoff.
\end{lemma}

\begin{proof}
Fix beliefs generated by a candidate strategy $r$. At posterior
confidence $p$, the risky--safe payoff difference is
\begin{equation}
 \Delta_r(p)
 =
 p[g+\gamma\mu_+(r)]
 +(1-p)[-\ell+\gamma\mu_-(r)+\tau]
 -\gamma\mu_S(r). \label{eq:deltaany}
\end{equation}
It is affine in $p$. Both recommendations occur in a nonpooling
equilibrium, so there is an indifference point in the interior of the
confidence support. If the slope of \eqref{eq:deltaany} were negative,
the best response would be a lower cutoff and, at $p=1$,
\[
 \Delta_r(1)=g+\gamma\mu_+(r)-\gamma\mu_S(r)<0.
\]
Yet Lemma~\ref{lem:arbitrary} and Assumption~\ref{ass:usd} imply that
this expression is strictly positive. If the slope were zero,
coexistence of the two recommendations would require
$\Delta_r(p)=0$ throughout the support, producing the same
contradiction. The slope is therefore positive, and the best response
is an upper cutoff. The confidence distribution is atomless, so mixing
at the unique indifference point is immaterial.
\end{proof}

\begin{theorem}\label{thm:nonpooling}
Suppose Assumptions~\ref{ass:interior} and~\ref{ass:usd} hold. Fix
$c\in(\ubar,\obar)$ with $D(c)>0$ and offer the least-cost contract
\[
 (t_S,t_+,t_-)=(0,0,\sigma(c)).
\]
Then the rule $R\Longleftrightarrow p(x)\geq c$ is the unique
nonpooling equilibrium among all measurable, possibly mixed advice
strategies.
\end{theorem}

\begin{proof}
Under beliefs generated by the target cutoff, the marginal payoff
difference is
\[
 \Delta_c(c)=(1-c)[\sigma(c)-\sigma(c)]=0.
\]
Its slope can be written, using marginal indifference, as
\[
 \frac{g+\gamma\mu_+(c)-\gamma\rho}{1-c}>0.
\]
Thus the target rule is sequentially rational.

Now consider any nonpooling equilibrium. By
Lemma~\ref{lem:endomon}, it is an upper cutoff $q$. Its marginal
indifference condition is
\[
 \sigma(c)=\sigma(q).
\]
Lemma~\ref{lem:monosigma} implies $q=c$.
\end{proof}

The preceding theorem is refinement free, and we prefer to keep it that way
rather than absorb a belief restriction into the statement. It does, however,
deliberately concern nonpooling equilibria only. At an all-safe assessment, outcome beliefs are
off path; unrestricted perfect Bayesian equilibrium permits arbitrary
beliefs that can sustain pooling. We therefore state the boundary
refinement used for the full result.

\begin{definition}
At all risk, beliefs are the continuous limits generated by upper
cutoffs $q\downarrow\ubar$. At all safety, they are the continuous
limits generated by upper cutoffs $q\uparrow\obar$. Hence all-risk
beliefs satisfy
\[
 \mu_S=\mu_+=\mu_-=\rho,
\]
whereas all-safe beliefs satisfy
\[
 \mu_S=\rho,\qquad
 \mu_+=\frac{\rho(1+h)}{1+b},\qquad
 \mu_-=\frac{\rho(1-h)}{1-b}.
\]
\end{definition}

\begin{theorem}\label{thm:full}
Under the assumptions of Theorem~\ref{thm:nonpooling} and
upper-tail-consistent boundary beliefs, the least-cost contract has a
unique equilibrium, up to behavior at the null cutoff signal: the
target upper cutoff $c$.
\end{theorem}

\begin{proof}
Theorem~\ref{thm:nonpooling} establishes uniqueness among nonpooling
strategies. At all risk, the lowest-confidence adviser's limiting
payoff difference is
\[
 (1-\ubar)[\sigma(c)-\sigma(\ubar)]<0.
\]
A positive-measure neighborhood of low signals therefore deviates to
safety. At all safety, the highest-confidence adviser's limiting
payoff difference is
\[
 (1-\obar)[\sigma(c)-\sigma(\obar)]>0,
\]
so a positive-measure neighborhood of high signals deviates to risk.
Both inequalities follow from the strict decrease of $\sigma$. Thus
neither pooling strategy is an equilibrium.
\end{proof}

\begin{remark}
Upper-tail consistency is an explicit monotone-deviation refinement,
not a consequence of Bayes' rule alone. Without it, the safe statement
is Theorem~\ref{thm:nonpooling}: unique implementation among all
nonpooling equilibria. An alternative is to impose a further
boundary-dominance inequality that rules out all-safe pooling under
every Bayes-consistent tremble:
\[
 \obar g-(1-\obar)\ell+(1-\obar)\sigma(c)
 >
 \gamma\frac{2\rho(1-\rho)(h-l)b}{1-b^2}.
\]
The numerical specification used below satisfies this inequality, but
only narrowly---the two sides differ by well under ten percent at the
benchmark parameters---and the refinement-based formulation is therefore the
more transparent one for the main text.\footnote{We are not indifferent
between the two routes. The dominance inequality is a restriction on
primitives and can be checked; the refinement is a restriction on beliefs and
cannot. The reason we lead with the refinement is that the inequality is
sufficient but far from necessary, so leading with it would make the pooling
claim look more parameter-dependent than it is.}
\end{remark}

\begin{corollary}\label{cor:lfunique}
Under Assumptions~\ref{ass:interior} and~\ref{ass:usd}, the zero-payment
game has a unique nonpooling equilibrium, the upper cutoff $\cLF$. It
is the unique equilibrium under upper-tail-consistent boundary beliefs.
\end{corollary}

\begin{proof}
By Proposition~\ref{prop:lf}, $\sigma(\cLF)=0$. Repeat the proofs of
Theorems~\ref{thm:nonpooling} and~\ref{thm:full} with $\tau=0$.
\end{proof}

\section{Capped failure protection}

The cap is the one place where success pay enters an optimal contract, and it
enters in a specific order: only after downside protection has been exhausted.
Suppose failure protection is subject to an exogenous cap
\[
 0\leq t_-\leq\bar t,
\]
while success pay is uncapped.

\begin{proposition}\label{prop:cap}
Fix an interior target $c$ with $D(c)>0$. The unique least-cost
incremental contract subject to the cap is
\begin{align}
 t_S^*(c;\bar t)&=0, \label{eq:captS}\\
 t_-^*(c;\bar t)&=\min\{\bar t,\sigma(c)\}, \label{eq:captm}\\
 t_+^*(c;\bar t)
 &=\frac{1-c}{c}[\sigma(c)-\bar t]_+. \label{eq:captp}
\end{align}
Thus failure protection is exhausted before success pay is used.
\end{proposition}

\begin{proof}
As in Theorem~\ref{thm:selection}, $t_S>0$ is both costly and
counterproductive. Put $y_+=ct_+$ and $y_-=(1-c)t_-$. The residual
problem is
\[
 \min_{y_+,y_-\geq0}
 \left\{
 \frac{P_+(c)}{c}y_+
 +\frac{P_-(c)}{1-c}y_-
 \right\}
\]
subject to
\[
 y_++y_-=D(c),\qquad y_-\leq(1-c)\bar t.
\]
Upper-tail selection makes the coefficient on $y_-$ strictly smaller.
The solution assigns as much incentive as the cap permits to $y_-$ and
the residual to $y_+$, yielding
\eqref{eq:captS}--\eqref{eq:captp}.\footnote{Writing the program in the
variables $y_+=ct_+$ and $y_-=(1-c)t_-$ is what makes the argument a
one-line linear program: the incentive constraint becomes an equality
budget in $(y_+,y_-)$, and the only remaining question is which of the two
unit prices is smaller. The prices are $P_+(c)/c$ and $P_-(c)/(1-c)$, and
the comparison is exactly \eqref{eq:costgap}.}
\end{proof}

\begin{proposition}\label{prop:capimpl}
Under Assumption~\ref{ass:usd}, the contract in
Proposition~\ref{prop:cap} uniquely implements the target among all
nonpooling equilibria. It fully implements the target under
upper-tail-consistent boundary beliefs.
\end{proposition}

\begin{proof}
At an alternative upper cutoff $q$, marginal indifference requires
\begin{equation}
 \sigma(q)
 =
 t_-+\frac{q}{1-q}t_+. \label{eq:capcross}
\end{equation}
The left side is strictly decreasing by
Lemma~\ref{lem:monosigma}; the right side is weakly increasing and is
strictly increasing when $t_+>0$. Since the target solves
\eqref{eq:capcross}, it is the unique upper-cutoff solution.

Under marginal indifference, the slope of the risky--safe payoff
difference is
\[
 \frac{g+\gamma\mu_+(q)+t_+-\gamma\mu_S(q)}{1-q}.
\]
Assumption~\ref{ass:usd} makes it positive under every feasible
selection, so Lemma~\ref{lem:endomon}'s unrestricted-strategy argument
continues to apply. This establishes uniqueness among nonpooling
strategies. Finally, the sign of the marginal deviation at either
pooling boundary is determined by the difference between the decreasing
left side and the increasing right side of \eqref{eq:capcross}; the
proof of Theorem~\ref{thm:full} therefore excludes both pooling
strategies under the same refinement.
\end{proof}

Let $C_{\bar t}(c)$ denote expected payments under the capped contract
and let
\[
 C_\infty(c)=P_-(c)\sigma(c).
\]
Define
\begin{equation}
 Q(c)
 =
 \frac{1-c}{c}P_+(c)-P_-(c)
 =
 P_R(c)\frac{\bar p_R(c)-c}{c}>0. \label{eq:Qdef}
\end{equation}

\begin{proposition}\label{prop:capcost}
Expected compensation can be written as
\begin{equation}
 C_{\bar t}(c)
 =
 C_\infty(c)+Q(c)[\sigma(c)-\bar t]_+. \label{eq:capcost}
\end{equation}
It falls strictly as $\bar t$ rises while the cap binds and is constant
once the cap is slack. On the encouragement region
$c\in[\cFB,\cLF)$, it is strictly decreasing in the target $c$.
\end{proposition}

\begin{proof}
Substituting \eqref{eq:captm}--\eqref{eq:captp} into expected payments
gives \eqref{eq:capcost}. At a fixed target,
\[
 \frac{\partial C_{\bar t}(c)}{\partial\bar t}
 =
 \begin{cases}
 -Q(c)<0,&\bar t<\sigma(c),\\
 0,&\bar t>\sigma(c).
 \end{cases}
\]
For target monotonicity, note that
\[
 P_-'(c)=-(1-c)f(c),\qquad
 C_\infty'(c)
 =-(1-c)f(c)\sigma(c)+P_-(c)\sigma'(c)<0.
\]
A direct cancellation gives
\[
 Q'(c)=-\frac{P_+(c)}{c^2}<0.
\]
When the cap binds, differentiating \eqref{eq:capcost} produces
\[
 C_{\bar t}'(c)
 =
 C_\infty'(c)
 +Q'(c)[\sigma(c)-\bar t]
 +Q(c)\sigma'(c)<0.
\]
When it is slack, $C_{\bar t}'=C_\infty'<0$.
\end{proof}

\begin{corollary}\label{cor:capcs}
The smallest and largest optimal cutoffs are weakly decreasing in
$\bar t$. If the optimum is unique,
\[
 \bar t_2>\bar t_1
 \quad\Longrightarrow\quad
 c^*(\bar t_2)\leq c^*(\bar t_1).
\]
\end{corollary}

\begin{proof}
For $\bar t_2>\bar t_1$, the gain from relaxing the cap is
\[
 Q(c)\left(
 [\sigma(c)-\bar t_1]_+
 -[\sigma(c)-\bar t_2]_+
 \right).
\]
It is weakly decreasing in $c$ because both $Q$ and $\sigma$ are
strictly decreasing. The objective therefore has decreasing
differences in $(c,\bar t)$. Standard one-dimensional monotone
comparative statics gives the stated ordering of the extremal
optimizers.
\end{proof}

\section{The organization's optimal cutoff}

This section adds the second margin. The organization now chooses the target
as well as the contract, and the two results that matter are that the target
never sits at first best and that, in the canonical uncapped problem, it is
unique for reasons traceable to primitives rather than to an assumption made
for convenience.

Let $F$ and $f$ denote the distribution and density of posterior
confidence. Gross project surplus under cutoff $c$ is
\begin{equation}
 B(c)=
 \int_c^{\obar}\{pG-(1-p)L\}\,dF(p), \label{eq:Bdef}
\end{equation}
so
\begin{equation}
 B'(c)=-(G+L)(c-\cFB)f(c). \label{eq:Bprime}
\end{equation}
The organization maximizes
\[
 W_{\bar t}(c)=B(c)-C_{\bar t}(c).
\]

\begin{theorem}\label{thm:location}
An optimal target exists, and every optimal target satisfies
\begin{equation}
 c^*\in(\cFB,\cLF]. \label{eq:location}
\end{equation}
Thus the organization either leaves laissez faire unchanged or corrects
the distortion only partially. Exact first-best correction is never
optimal.
\end{theorem}

\begin{proof}
Continuity and compactness give existence. Any $c<\cFB$ includes
negative-value projects that $c=\cFB$ excludes and, by
Proposition~\ref{prop:capcost}, costs strictly more to implement. It is
therefore dominated by $\cFB$. Any $c>\cLF$ excludes positive-value
projects relative to laissez faire and cannot cost less than the
zero-payment implementation of $\cLF$; it is dominated by $\cLF$.
Hence every optimum belongs to $[\cFB,\cLF]$.

At first best, $B'(\cFB)=0$ while
$C_{\bar t}'(\cFB+)<0$. Consequently,
\[
 W_{\bar t}'(\cFB+)=-C_{\bar t}'(\cFB+)>0,
\]
and $\cFB$ itself cannot be optimal.
\end{proof}

In the uncapped canonical problem, strict concavity follows from a
one-dimensional primitive condition. Define
\[
 \kappa\equiv\gamma\rho(1-\rho)(h-l),\qquad
 s\equiv g+\ell,\qquad
 z^{\mathrm{FB}}\equiv\frac{\cFB-\ubar}{b},
\]
and
\begin{equation}
 \Phi(z)\equiv
 \frac{bz(1-z)^2}
 {(1+b-2bz)(1+bz)}. \label{eq:Fcurvature}
\end{equation}
Primes on \(\Phi\) in the next result denote derivatives with respect to
the raw cutoff \(z\). The symbol \(\Phi\) is used here rather than \(F\)
because \(F\) already denotes the confidence distribution, and \(s\) rather
than \(r\) because \(r\) already denotes an advice rule.

\begin{proposition}
\label{prop:primitivecurvature}
In the canonical model without a cap, the primitive inequality
\begin{equation}
 sb+\kappa\Phi''(z^{\mathrm{FB}})>0 \tag{PC}\label{eq:pc}
\end{equation}
is sufficient for \(C_\infty\) to be strictly convex on
\([\cFB,\cLF]\). In particular, Assumption~\ref{ass:usd} and
\(\cFB\geq1/2\) imply \eqref{eq:pc}. Under either formulation,
\(W_\infty\) is strictly concave on \([\cFB,\cLF]\).
\end{proposition}

\begin{proof}
Write \(c(z)=\ubar+bz\) and define
\[
 H(z)\equiv\frac{P_-(z)}{1-c(z)}
 =\frac{(1-z)(1-bz)}{1+b-2bz},
 \qquad
 J(z)\equiv[c(z)-\cFB]H(z).
\]
Equations~\eqref{eq:Kclosed} and \eqref{eq:sigmadef} give the
decomposition
\begin{equation}
 C_\infty(z)=\kappa\Phi(z)-sJ(z). \label{eq:Cdecomp}
\end{equation}
Direct differentiation yields
\begin{align}
 H'(z)
 &=-\frac12-\frac{(1-b)^2}{2(1+b-2bz)^2}
 \leq-\frac12, \label{eq:Hprime}\\
 H''(z)
 &=-\frac{2b(1-b)^2}{(1+b-2bz)^3}\leq0. \label{eq:Hsecond}
\end{align}
For \(z\geq z^{\mathrm{FB}}\), therefore,
\begin{equation}
 J''(z)
 =2bH'(z)+[c(z)-\cFB]H''(z)\leq-b. \label{eq:Jbound}
\end{equation}
The other component satisfies
\begin{align}
 \Phi''(z)
 &=
 \frac{2(1-b)^2(1+b)}
 {(b+3)(1+b-2bz)^3}
 -
 \frac{2(1+b)^2}
 {(b+3)(1+bz)^3}, \label{eq:Fsecond}\\
 \Phi'''(z)
 &=
 \frac{12b(1-b)^2(1+b)}
 {(b+3)(1+b-2bz)^4}
 +
 \frac{6b(1+b)^2}
 {(b+3)(1+bz)^4}>0. \label{eq:Fthird}
\end{align}
Thus \(\Phi''(z)\geq\Phi''(z^{\mathrm{FB}})\) on the target interval.
Combining \eqref{eq:Cdecomp}--\eqref{eq:Fthird} gives
\[
 \frac{d^2 C_\infty}{dz^2}
 =\kappa\Phi''(z)-sJ''(z)
 \geq
 \kappa\Phi''(z^{\mathrm{FB}})+sb.
\]
Condition~\eqref{eq:pc} proves strict convexity. Because \(c(z)\) is a
positive affine transformation, strict convexity also holds with respect
to \(c\).

It remains to establish the primitive specialization. If
\(\cFB\geq1/2\), then \(z^{\mathrm{FB}}\geq1/2\). By
\eqref{eq:Fthird},
\[
 \Phi''(z^{\mathrm{FB}})\geq\Phi''(1/2),
 \qquad
 \Phi''(1/2)+b
 =
 \frac{b^3(2b^2+5b+4)}{(b+2)^3}>0.
\]
Proportional alignment gives \(g=s(1-\cFB)\), while
Assumption~\ref{ass:usd} gives \(g>\kappa/(1-b)\). In particular,
\(s>\kappa\). Hence
\[
 \kappa\Phi''(z^{\mathrm{FB}})+sb
 >
 -\kappa b+sb
 =b(s-\kappa)>0.
\]
Finally, in the canonical model
\[
 B''(c)=-\frac{G+L}{b}<0.
\]
Strict convexity of \(C_\infty\) therefore makes
\(W_\infty=B-C_\infty\) strictly concave.
\end{proof}

Condition~\eqref{eq:pc} can hold when \(\cFB<1/2\); the loss-dominance
condition in the main text is a transparent sufficient specialization, not a
boundary of the result.\footnote{The step \(\Phi''\geq-b\) is where the
specialization is spent, and it is generous: it discards the entire first term
of \eqref{eq:Fsecond}. What \eqref{eq:pc} needs is only that the convexity
contributed by the linear surplus term, \(sb\), not be overturned by the
curvature of the wedge at the first-best cutoff, and since \(s>\kappa\) under
Assumption~\ref{ass:usd} this leaves considerable room.} A binding cap or a noncanonical
confidence distribution requires an additional curvature condition.
The following assumption states that condition directly.

\begin{assumption}\label{ass:sc}
On $[\cFB,\cLF]$, the marginal project loss
\[
 \Psi(c)=(G+L)(c-\cFB)f(c)
\]
is strictly increasing, while the marginal compensation saving
\[
 M_{\bar t}(c)=-C_{\bar t}'(c)
\]
is weakly decreasing, with one-sided derivatives at a cap kink.
\end{assumption}

A primitive sufficient condition for the first part is
\[
 f(c)+(c-\cFB)f'(c)>0.
\]
In the canonical model $f(c)=1/b$, so this part holds automatically.
The second part is a transparent curvature restriction on implementation
cost. Equivalently, one may assume directly that $W_{\bar t}$ is
strictly concave.

\begin{theorem}\label{thm:optimum}
Suppose either that the problem is uncapped,
Assumption~\ref{ass:usd} holds, and \(\cFB\geq1/2\), or that
Assumption~\ref{ass:sc} holds. The optimal target is unique. The
organization intervenes if and only if
\begin{equation}
 (G+L)(\cLF-\cFB)f(\cLF)
 >
 M_{\bar t}(\cLF-). \label{eq:intgeneral}
\end{equation}
If the inequality holds, the unique optimum is in
$(\cFB,\cLF)$ and solves
\[
 (G+L)(c^*-\cFB)f(c^*)=M_{\bar t}(c^*).
\]
If it fails weakly, $c^*=\cLF$.
\end{theorem}

\begin{proof}
The objective derivative is
\[
 W_{\bar t}'(c)=M_{\bar t}(c)-\Psi(c).
\]
Under Proposition~\ref{prop:primitivecurvature},
\(M_\infty=-C_\infty'\) is strictly decreasing and, because
\(f=1/b\), \(\Psi\) is strictly increasing. Assumption~\ref{ass:sc}
imposes the same two properties directly in the other cases. Thus
\(W_{\bar t}'\) is strictly decreasing and is positive at $\cFB$. If
\eqref{eq:intgeneral} holds, it is negative at the right endpoint and
hence has exactly one interior zero. If the inequality fails, it remains
nonnegative throughout and the objective is maximized at $\cLF$.
\end{proof}

\begin{corollary}\label{cor:inttest}
Under the conditions of Theorem~\ref{thm:optimum}, every positive finite
cap is slack in a left neighborhood of $\cLF$; the same formula applies
without a cap. In either case, \eqref{eq:intgeneral} becomes
\begin{equation}
 (G+L)(\cLF-\cFB)f(\cLF)
 >
 -P_-(\cLF)\sigma'(\cLF). \label{eq:inttest}
\end{equation}
If failure protection is prohibited, $\bar t=0$, the right side is
instead
\[
 -\frac{1-\cLF}{\cLF}P_+(\cLF)\sigma'(\cLF),
\]
which is strictly larger by upper-tail selection.
\end{corollary}

\begin{proof}
Because $\sigma(\cLF)=0$, every positive cap is locally slack at
$\cLF$. Differentiating $C_\infty=P_-\sigma$ at the endpoint gives
\[
 M_{\bar t}(\cLF-)=-P_-(\cLF)\sigma'(\cLF).
\]
When $\bar t=0$, all incentive is supplied by success pay and
\[
 C_0(c)=P_+(c)\frac{1-c}{c}\sigma(c).
\]
Differentiation at $\cLF$ yields the alternative expression. Their
difference is $-\sigma'(\cLF)Q(\cLF)>0$.
\end{proof}

\begin{remark}
Outside the uncapped primitive case of
Proposition~\ref{prop:primitivecurvature}, dropping
Assumption~\ref{ass:sc} leaves Theorem~\ref{thm:location} valid but does
not guarantee uniqueness. Likewise, \eqref{eq:inttest} is then only a
local test for whether a small movement away from laissez faire is
profitable; it is not an if-and-only-if test for globally optimal
intervention.
\end{remark}

\section{Internal review and disclosure}

The disclosure result is the one place in the paper where an informational
instrument substitutes for a monetary one, and its logic is entirely about
what a veto reveals. Under transparency a veto and a failed launch are
informationally identical, so screening cannot touch the career wedge at all;
under confidentiality a veto is pooled with safe advice, and the pooled
history is strictly better news about ability than a failure is. Everything
else in this section follows from that observation.

After risky advice, an internal review detects a bad project with
probability $\lambda\in[0,1]$ and vetoes it. Good projects are never
vetoed. Conditional on a bad project and risky advice, detection is
independent of ability and the precise signal. A veto yields zero
material payoff for the adviser and the organization.

Under \emph{transparent review}, the market observes
$\{S,V,+,-\}$. Under \emph{confidential review}, it observes
$\{N,+,-\}$, where $N$ pools safe advice with an internal veto. The
employer privately observes the underlying event, can contract on it,
and keeps compensation private. Thus a veto payment does not unravel
confidentiality.

This section fixes an interior upper cutoff $z$, writes
\[
 q=p(z),\qquad
 \bar q=\bar p_R(z),
\]
and permits any differentiable increasing career-value function
$V(\mu)$. The linear specification used above is a special case.

\begin{lemma}\label{lem:reviewpost}
Under transparent review,
\[
 \mu_V(z,\lambda)=\mu_-(z).
\]
Under confidential review, define
\[
 m_\theta(z,\lambda)
 =
 z+\frac{\lambda}{2}(1-z)(1-a_\theta z).
\]
The no-launch posterior is
\begin{equation}
 \mu_N(z,\lambda)
 =
 \frac{\rho m_H(z,\lambda)}
 {\rho m_H(z,\lambda)+(1-\rho)m_L(z,\lambda)}. \label{eq:muN}
\end{equation}
For $\lambda>0$,
\[
 \mu_-(z)<\mu_N(z,\lambda)<\rho,
\]
and
\begin{equation}
 \frac{\partial\mu_N}{\partial\lambda}
 =
 -\frac{\rho(1-\rho)(h-l)z^2(1-z)}
 {2[\rho m_H+(1-\rho)m_L]^2}<0. \label{eq:muNprime}
\end{equation}
\end{lemma}

\begin{proof}
For type $\theta$, a veto and an implemented failure have likelihoods
\[
 \frac{\lambda}{2}(1-z)(1-a_\theta z)
 \quad\text{and}\quad
 \frac{1-\lambda}{2}(1-z)(1-a_\theta z).
\]
They differ by a type-independent factor, so Bayes' rule gives
$\mu_V=\mu_-$.\footnote{This is the substantive content of one-sided,
ability-independent detection. If the reviewer were better at catching the bad
projects of low-ability advisers, or occasionally vetoed good projects, a veto
would carry its own information about ability and $\mu_V$ would separate from
$\mu_-$; transparent review would then move the wedge, in a direction depending
on the correlation. We do not pursue that case, but it is where the assumption
binds.} Safe advice has probability $z$ for either type.
Adding its likelihood to the veto likelihood gives $m_\theta$ and hence
\eqref{eq:muN}. Equivalently, $\mu_N$ is a strict posterior-weighted
average of $\rho$ and $\mu_-$ when $\lambda>0$, proving the ranking.
Differentiating \eqref{eq:muN} yields \eqref{eq:muNprime}.
\end{proof}

Let
\[
 v_+=V(\mu_+),\qquad
 v_-=V(\mu_-),\qquad
 v_S=V(\rho),\qquad
 v_N(\lambda)=V(\mu_N(\lambda)).
\]
The marginal career wedges are
\begin{align}
 K_T(z,\lambda)
 &=
 v_S-\{qv_++(1-q)v_-\}, \label{eq:KT}\\
 K_C(z,\lambda)
 &=
 v_N-
 \left[
 qv_+
 +(1-q)\{(1-\lambda)v_-+\lambda v_N\}
 \right]. \label{eq:KC}
\end{align}

\begin{proposition}\label{prop:shield}
Transparent review leaves the career wedge unchanged:
\[
 \frac{\partial K_T}{\partial\lambda}=0.
\]
Confidential review strictly reduces it:
\[
 \frac{\partial K_C}{\partial\lambda}<0.
\]
Therefore, for every $\lambda>0$,
\[
 K_C(z,\lambda)<K_T(z,\lambda),
\]
and $K_T-K_C$ is strictly increasing in review precision.
\end{proposition}

\begin{proof}
Equation~\eqref{eq:KT} is independent of $\lambda$ because a veto and a
failure convey the same information under transparency. Rewrite
\eqref{eq:KC} as
\[
 K_C=
 \{1-(1-q)\lambda\}v_N
 -qv_+-(1-q)(1-\lambda)v_-.
\]
Its derivative is
\[
 \frac{\partial K_C}{\partial\lambda}
 =
 \{1-(1-q)\lambda\}
 V'(\mu_N)\frac{\partial\mu_N}{\partial\lambda}
 -(1-q)(v_N-v_-).
\]
The first term is strictly negative by \eqref{eq:muNprime} at every interior
$z$, and the second is strictly negative because $\mu_N>\mu_-$ implies
$v_N>v_-$. At $\lambda=0$, $\mu_N=\rho$ and $K_C=K_T$. The strict derivative
ranking proves all claims.
\end{proof}

The adviser's material payoff from risky advice at the margin is
\[
 qg-(1-q)(1-\lambda)\ell.
\]
For regime $j\in\{T,C\}$, define the total encouragement deficit
\begin{equation}
 D_j(z,\lambda)
 =
 K_j(z,\lambda)
 -\{qg-(1-q)(1-\lambda)\ell\}. \label{eq:Dreview}
\end{equation}

\begin{corollary}[Review reduces the deficit]\label{cor:reviewdeficit}
\begin{align}
 \frac{\partial D_T}{\partial\lambda}
 &=-(1-q)\ell<0,\\
 \frac{\partial D_C}{\partial\lambda}
 &=
 \frac{\partial K_C}{\partial\lambda}
 -(1-q)\ell
 <
 \frac{\partial D_T}{\partial\lambda}<0.
\end{align}
Moreover,
\[
 D_T-D_C=K_T-K_C>0
\]
for every $\lambda>0$, and the difference is increasing in $\lambda$.
\end{corollary}

\begin{proof}
Differentiate \eqref{eq:Dreview} and apply
Proposition~\ref{prop:shield}. The material loss-prevention term is
identical under the two disclosure regimes.
\end{proof}

Let $t_V\geq0$ be a privately observed payment after an internal veto.
Marginal indifference requires
\begin{equation}
 qt_+
 +(1-q)\{(1-\lambda)t_-+\lambda t_V\}
 -t_S
 =
 D_j(z,\lambda). \label{eq:reviewIC}
\end{equation}

\begin{theorem}\label{thm:reviewcost}
Fix $(z,\lambda)$ and suppose $D_j(z,\lambda)>0$. Every least-cost
schedule sets
\[
 t_S=t_+=0
\]
and satisfies
\begin{equation}
 (1-\lambda)t_-+\lambda t_V
 =
 \frac{D_j(z,\lambda)}{1-q}. \label{eq:downside}
\end{equation}
All such schedules have expected cost
\begin{equation}
 C_j^I(z,\lambda)
 =
 A(z)D_j(z,\lambda),\qquad
 A(z)
 =
 (1-z)\frac{1-\bar q}{1-q}. \label{eq:reviewcost}
\end{equation}
While the deficit remains positive, review strictly lowers the cost of
supporting the target. If $D_C>0$---and hence also $D_T>0$---then
\begin{equation}
 C_C^I(z,\lambda)<C_T^I(z,\lambda)
 \quad\text{for }\lambda>0, \label{eq:reviewranking}
\end{equation}
and the cost advantage of confidentiality is strictly increasing in
$\lambda$.
\end{theorem}

\begin{proof}
A safe payment works against \eqref{eq:reviewIC}. The cost per unit of
marginal incentive supplied by success pay is
\[
 (1-z)\frac{\bar q}{q}.
\]
The corresponding costs for failure and veto protection are both
\[
 (1-z)\frac{1-\bar q}{1-q}.
\]
Because $\bar q>q$, either downside instrument is strictly cheaper than
success pay.\footnote{Note that failure protection and veto protection carry
the \emph{same} unit price, because the factors $1-\lambda$ and $\lambda$ enter
the expected payment and the marginal incentive identically. The least-cost
schedule is therefore not unique in $(t_-,t_V)$; only the aggregate
\eqref{eq:downside} is pinned down. An organization that finds a veto payment
easier to justify than a failure payment may accordingly load the whole
transfer onto $t_V$ at no cost, which is a small but practically relevant
degree of freedom.} The least-cost schedules therefore use only failure and
veto protection and satisfy \eqref{eq:downside}. Their common expected
cost is \eqref{eq:reviewcost}. Since $A(z)>0$ is independent of
$\lambda$, Corollary~\ref{cor:reviewdeficit} proves the remaining
claims.
\end{proof}

\begin{corollary}\label{cor:reviewchoice}
Let the feasible review interval be $[0,\bar\lambda]$ and suppose
$D_C(z,\lambda)>0$ throughout it. Let review cost
$\Gamma(\lambda)$ and gross project surplus be the same under both
disclosure policies. Then the payoff advantage of confidentiality is
positive and strictly increasing in $\lambda$. Every optimal review
precision under confidentiality is weakly greater than every optimal
precision under transparency.
\end{corollary}

\begin{proof}
Gross project surplus and review cost cancel in the regime comparison,
leaving
\[
 \Pi_C(z,\lambda)-\Pi_T(z,\lambda)
 =
 A(z)\{K_T(z,\lambda)-K_C(z,\lambda)\},
\]
which is strictly increasing by Proposition~\ref{prop:shield}. If a
confidential-regime maximizer were below a transparent-regime
maximizer, moving the confidential policy to the latter point would
weakly increase the transparent payoff and strictly increase the
confidentiality advantage, contradicting optimality.
\end{proof}

\begin{remark}
The positive-deficit restriction is substantive. If review becomes
strong enough to make $D_j<0$, the fixed target requires a safe payment
to restrain risky advice. Since $D_C<D_T$, confidentiality can then be
\emph{more} expensive. Thus neither
\eqref{eq:reviewranking} nor Corollary~\ref{cor:reviewchoice} is claimed
globally on $[0,1]$. One-sided detection, type-independent review,
credible secrecy, and private contractibility are also essential.
\end{remark}

\begin{remark}
Theorem~\ref{thm:reviewcost} is a fixed-target support result. Define
\[
 \sigma_j(c,\lambda)=\frac{D_j(c,\lambda)}{1-c}.
\]
On any compact cutoff interval, $\sigma_j'$ is continuous in
$(c,\lambda)$ and equals the strictly negative baseline derivative at
$\lambda=0$. Hence, for sufficiently small review precision, it
remains negative and the target is unique among regular upper-cutoff
candidates. A global claim over arbitrary strategies and pooling
assessments would require extending the equilibrium-selection argument
of Section~5 and is not needed for the disclosure-cost theorem.
\end{remark}

\section{Numerical witness}

The following numerical exercise is a consistency check, not evidence: its
purpose is to confirm that the assumptions are mutually satisfiable and that
the objects in the theorems take sensible values at a point where every
hypothesis holds. The benchmark uses
\[
 h=.8,\quad l=.2,\quad \rho=\frac12,\quad\gamma=1,
 \quad g=\ell=.4,\quad G=L=10.
\]
The uniform success-dominance bound equals $.3<g$. The verified
solution is
\[
 z^{\mathrm{FB}}=.500000,\qquad
 z^0=.550202,\qquad
 z^*=.515377,
\]
with
\[
 p(z^*)=.507689,\qquad
 \bar p_R(z^*)=.628844.
\]
At the optimum,
\[
 D(z^*)=.013914,\qquad
 t_-^*(z^*)=.028262.
\]
Expected failure-protection cost is $.005083$, compared with $.008352$
for success-only pay, a reduction of approximately $39$ percent.

At the same target, the confidential-review deficit reaches zero near
$\lambda=.0492$, while the transparent-review deficit reaches zero near
$\lambda=.0707$. The disclosure ranking in
Theorem~\ref{thm:reviewcost} is therefore illustrated on
$\lambda\in[0,.04]$. The associated verification script checks all
closed forms, equilibrium inequalities, cost rankings, and this
positive-deficit domain.\footnote{The benchmark sits at $\cFB=1/2$ exactly, so
Proposition~\ref{prop:primitivecurvature} applies at its boundary and the
uniqueness of $z^*$ is not being obtained from a comfortable interior case. It
also satisfies Assumption~\ref{ass:usd} with slack ($0.3<0.4$) and the
boundary-dominance inequality of Section~5 only narrowly, which is why the
refinement rather than that inequality is used in the statement of full
uniqueness.}

\end{document}